\documentclass{aa}  
\usepackage{graphicx}
\def\ocen{$\omega$Cen}
\def\nafe{[Na/Fe]}

\def\ofe{[O/Fe]}
\def\nfe{[N/Fe]}
\def\cfe{[C/Fe]}
\def\feh{[Fe/H]}
\def\alfe{[Al/Fe]}
\def\mgfe{[Mg/Fe]}
\def\ne22{$^{22}$Ne}
\def\c12{$^{12}$C}
\def\n14{$^{14}$N}
\def\o16{$^{16}$O}
\def\tbce{T_{\rm bce}}
\def\msun{$M_{\odot}$}

\begin{document}
\title{Massive AGB models of low metallicity: the implications for
  the self-enrichment scenario in metal poor Globular Clusters}
\authorrunning {Ventura \& D'Antona}
\titlerunning {AGB models of low metallicity}


   \author{P. Ventura \and F. D'Antona}

   \offprints{P. Ventura}

   \institute{INAF - Observatory of Rome, Via Frascati 33,
              00040 MontePorzio Catone (RM) - Italy
              \email{ventura, dantona@oa-roma.inaf.it}
             }


 
  \abstract
   {We present the physical and chemical properties of intermediate-mass stars
    models of low metallicity, evolved along the thermal pulse phase.}
   {The target of this work is to extend to low metallicities, Z=1,2 and 
    $6\times 10^{-4}$, the models
    previously computed for chemistries typical of Globular Clusters of 
    an intermediate metallicity (Z=0.001), and for the most metal-rich clusters
    found in our Galaxy (Z=0.004); the main goal is to test the self-enrichment 
    scenario also for metal poor Globular Clusters}
   {We calculated three grids of intermediate-mass models with metallicities 
    $Z=10^{-4}$, $2\times 10^{-4}$, and $6\times 10^{-4}$; the evolutionary 
    sequences are followed from the pre-main sequence throughout the AGB 
    phase, almost until the ejection of the whole envelope. We discuss the
    chemistry of the ejecta, and in particular the mass fractions of those
    elements that have been investigated during the many, deep, spectrocopic
    surveys of Globular Clusters}
   {Although the data for oxygen and sodium are scarce for low metallicity Globular Clusters,
   the few data for the unevolved stars in NGC~6397 are compatible with the models.
   Further, we find good agreement with the C--N anticorrelation of unevolved stars in 
   the cluster M15. In this cluster, however, no stars having low oxygen ([O/Fe]$\sim -1$)
   have been detected. The most massive, very metal poor clusters, should contain
   such stars, according to the present models. At the lowest
   metallicity $Z=10^{-4}$, the ejecta of the most massive AGBs have C/O$>1$, due to
   the dramatic decrease of the oxygen abundance. We discuss the possible implications
   of this prediction.}
   {}


   \maketitle
%

\section{Introduction}
In the last decades, the evolution of massive Asymptotic Giant 
Branch stars (AGB, i.e. stars with
masses $4M_{\odot} \leq M \leq 7M_{\odot}$ during the thermal 
pulses phase) has been the subject
of extended investigations by several research groups, as they were suggested 
to be the main responsible for the star-to-star differences in the
surface chemistry of Globular Clusters (GC) stars.
GC stars exhibit considerable
differences in the surface abundances of the ``light'' elements (A$<$30)
(Kraft 1994), showing well defined abundance patterns, that involve all 
the species up to aluminum (see Carretta 2006 for a review on this topic).
The idea that these apparent anomalies were formed ``in situ'' by some non
canonical extra-mixing from the bottom of the convective envelope during
the Red Giant Branch (RGB) evolution (Denissenkov \& Weiss 2001) 
was frustrated by the discovery that even cool structures like Turn-Off (TO) 
or SubGiant Branch (SGB) stars showed the same patterns (Gratton et al. 2001), 
thus indicating that the current observed surface chemical compositions were 
essentially the same with which the stars formed. This opened the way to  
``self-enrichment'' scenarios: the stars with the anomalous chemistry (second 
generation ---hereinafter SG--- stars) formed in an interstellar medium polluted 
by the winds of an earlier generation of stars (the first generation or FG).
Either the winds from massive AGBs (Cottrell \& Da Costa 1981; Ventura et al. 2001) 
or those from fast rotating massive stars (Prantzos \& Charbonnel 2006) have been 
proposed as progenitors of the SG; the latter scenario is described in detail in 
Decressin et al. (2007) and references therein;  here we follow the first 
hypothesis. The massive AGBs evolve at  very short ages ($\sim 40-100$Myr) 
compared to the typical ages of GCs ($\sim 10-15$Gyr) and achieve, at the
bottom of their surface convective zone, a very advanced nucleosynthesis
(Hot Bottom Burning, hereinafter HBB, Bl\"ocker \& Sch\"onberner 1991;
D'Antona \& Mazzitelli 1996), the products of which would be carried to the 
surface by the convective currents (Ventura et al. 2001): the medium in which
this generation of stars evolved would thus show the signature of such 
nucleosynthesis, because the low velocities ($\sim 10-20$Km/s) of these winds 
allow them to be kept inside the cluster.
If the Initial Mass Function (IMF) of the FG is standard (e.g. a Salpeter (1955)
or Kroupa et al. (1993) IMF), the gas contained in 
the massive AGB ejecta is too scarce to give origin to a SG containing the large fraction of
today's cluster stars shown from observation. Therefore, it has been suggested that
most of FG stars have been lost, as confirmed by some
dynamical and N--body models (D'Ercole et al. 2008) or that the GC has collected the 
AGB gas from a much larger environment, e.g. if it was born in a dwarf galaxy that 
today is dispersed (Bekki \& Norris 2006, Renzini 2008).
A robust prediction from stellar evolution models is that the ejecta from 
massive AGBs, like those from massive stars (Smith 2006), are expected to be helium-rich. 
The existence of a helium-rich population
($Y>0.30$) provides an appealing explanation for the existence of some GCs 
with Horizontal Branches (HBs) showing an extended blue tail (D'Antona et
al. 2002). This idea has been elaborated  in many
subsequent works (e.g. D'Antona \& Caloi 2004, Lee et al. 2005, Caloi \& D'Antona 2005,
2007, Busso et al. 2007, D'Antona \& Caloi 2008). The presence of a fraction of stars
enriched in helium has been inferred also by the presence of a blue main
sequence in \ocen\  (Bedin et al. 2004; Norris 2004; Piotto et al. 2005), 
and in NGC 2808 (Piotto et al. 2007).

However, the debate concerning the possible role that massive AGBs may 
have played in the self-enrichment scenario is still open, because AGB
modelling proves to be one of the most delicate and uncertain tasks in
the whole context of stellar evolution. This is the reason why results
presented by the various groups working on this topic are sometimes
extremely different, leading to opposite conclusions concerning
many physical and chemical properties of the evolution of this class
of objects (Denissenkov \& Herwig 2003; Fenner at al. 2004; Karakas 
\& Lattanzio 2007, KL07; Ventura et al. 2001).

Ventura \& D'Antona (2005a,b; 2006) showed that the discrepancies in the 
results obtained by the various investigators are a mere consequence of 
the different description which they make of some physical phenomena.
More in details, Ventura \& D'Antona (2005a) showed that the efficiency
of the convective model adopted may change substantially the evolution
of the main physical properties of these stars, e.g. the duration of
the whole AGB phase, the maximum luminosity reached, and the degree
of nucleosynthesis achieved at the bottom of the convective envelope.
The main two arguments against the self-enrichment scenario raised by
some investigators, namely that the winds of even the most massive AGBs
should show only a modest extent of oxygen depletion and a great
enhancement of the overall CNO abundances (Fenner at al. 2004), at 
odds with the observational evidence (Ivans et al. 1999), are entirely 
due to the use of the Mixing Length Theory (MLT, Vitense 1953); when 
the Full Spectrum of Turbulence (FST, Canuto \& Mazzitelli 1991) approach 
is used to model convection, a much stronger HBB and a much smaller CNO 
enhancement is found (Ventura \& D'Antona 2005a).

The uncertainties associated with mass loss prove to have a smaller
impact on the results obtained, as extensively discussed in Ventura
\& D'Antona (2005b).

Ventura \& D'Antona (2008a,b) presented their most recent and updated
AGB models to test the self-enrichment scenario in the cases of GCs of 
intermediate (Z=0.001) and higher (Z=0.004) metallicities.
These works were a more systematic treatment of this issue, aimed at 
integrating their qualitative approach discussed in D'Antona \& Ventura
(2007), limited only to the Z=0.001 case.
The main results of these investigations were that the most massive 
FST AGB models ($M\sim 5-6M_{\odot}$) produce ejecta whose chemical 
composition is in agreement with the abundance patterns observed in 
TO and SGB stars in GCs, in terms of oxygen-sodium, oxygen-aluminum, 
and carbon-nitrogen anticorrelations. 
These results require a choice of the nuclear cross-sections 
involving proton captures by $^{22}$Ne, $^{23}$Na, $^{25}$Mg and
$^{26}$Mg nuclei, within the range allowed by the uncertainties (Hale et 
al 2002; 2004). The most extreme anomalies, observed only in low gravity 
giants, can be explained on the basis of a possible non canonical extra-mixing 
during the RGB phase, that could be favoured by the lower height of the 
entropy barrier at the hydrogen-helium interface expected in the stars
belonging to the SG (D'Antona \& Ventura 2007).

The aim of this work is to extend to lower metallicities the results of
the above mentioned investigations. To this scope we calculated new evolutionary
sequences for models of intermediate mass with metallicities  
$Z=10^{-4}$, $2\times 10^{-4}$, $6\times 10^{-4}$. These computations complete
the theoretical framework concerning massive AGBs with chemistry typical of
GC stars, and allow to begin to test the validity of the self-enrichment scenario
in the case of the most metal-poor GCs, comparing the chemistry of the 
theoretical ejecta of these models with the abundance patterns shown
by the spectroscopic investigations of the GCs M15 (Sneden et al. 1997,
Cohen et al. 2005) and NGC~6397 (Carretta et al. 2005).

The paper is organized as follows. Sect.2 describes the physical and chemical
ingredients used to calculate the evolutionary sequences. The physical properties 
of the models are presented in Sect.3. The chemical content of their
ejecta, and its dependence on mass and metallicity, is discussed in Sect.4,
and compared with the observations in Sect.5.


\section{The physical and chemical inputs}
All the evolutions presented in this work have been calculated by
means of the ATON code for stellar evolution, with the numerical 
structure described in details in Ventura et al. (1998).
We adopt the latest opacities by Ferguson et al. (2005) at temperatures 
lower than 10000 K and the OPAL opacities in the version documented by Iglesias 
\& Rogers (1996). The mixture adopted is alpha-enhanced, with $[\alpha$/Fe$]=0.4$
(Grevesse \& Sauval 1998). The conductive opacities are taken 
from Poteckhin (2006, see the web page www.ioffe.rssi.ru/astro/conduct/), 
and are harmonically added to the radiative opacities.
Tables of the equation of state are generated in the (gas) pressure-temperature
plane, according to the latest release of the OPAL EOS (2005), overwritten in
the pressure ionization regime by the EOS by Saumon, Chabrier \& Van Horn (1995),
and extended to the high-density, high-temperature domain according to the
treatment by Stoltzmann \& Bl\"ocker (2000). 
Convection was modelled according to the FST prescription.
Mixing of chemicals within convective zones has been treated as a
diffusive process. We follow the approach by Cloutman \& Eoll (1976), solving 
for each chemical species the diffusive-like equation:
$$
{dX_i\over dt}=\big( {\partial X_i\over \partial t} \big)_{nucl}+
{\partial \over \partial m_r} \big[ (4\pi r^2\rho )^2D{\partial X_i \over \partial m_r} \big]
\eqno{(1)}
$$
where D is the diffusion coefficient, for which, given the convective velocity
$v$ and the scale of mixing $l$, a local approximation ($D\sim {1\over 3}vl$) 
is adopted. The borders of the convective regions are fixed according to the
Schwarzschild criterium. We considered extra-mixing from all the formal
convective boundaries up to the beginning of the AGB phase: convective 
velocities are assumed to decay exponentially with an e-folding distance 
described by the free-parameter $\zeta$, that was set to $\zeta=0.02$,
according to the calibration provided in Ventura et al. (1998), where the
interested reader can also find a complete discussion regarding the variation 
of the convective velocities in the proximities of the convective borders.
No extra-mixing was assumed during the whole AGB phase: these results
provide therefore a conservative estimate of the extent of the Third Dredge-up
following each thermal pulse.

Mass loss was described according to the Bl\"ocker (1995) formulation, that
is more accurate than the basic Reimer's recipe to describe the steep 
increase of mass loss with luminosity as the stars climb the AGB on the
HR diagram. The full expression is
$$
\dot M=4.83 \times 10^{-22} \eta_R M^{-3.1}L^{3.7}R
\eqno{(2)}
$$
where $\eta_R$ is the free parameter entering the Reimers' prescription,
for which we used $\eta_R=0.02$, according to the calibration based on
the luminosity function of lithium rich stars in the Magellanic Clouds 
given in Ventura et al.(2000).
The nuclear network includes 30 elements (up to $^{31}$P) and 64
reactions, a full list of which can be found in Ventura \& D'Antona (2005a).
The relevant cross sections are taken from the recommended values
pf the NACRE compilation (Angulo et al. 1999), with only the following exceptions:

\begin{enumerate}

\item{
$^{14}$N(p,$\gamma$)$^{15}$O (Formicola et al. 2004)
}

\item{
$^{22}$Ne(p,$\gamma$)$^{23}$Na (Hale et al. 2002)
}

\item{
$^{23}$Na(p,$\gamma$)$^{24}$Mg (Hale et al. 2004)
}
\item{
$^{23}$Na(p,$\alpha$)$^{20}$Ne (Hale et al. 2004)
}

\item{
$^{25}$Mg(p,$\gamma$)$^{26}$Al (NACRE, upper limits)
}
\item{
$^{26}$Mg(p,$\gamma$)$^{27}$Al (NACRE, upper limits)
}

\end{enumerate}

\section{The physical properties of the low Z AGB models}
The models presented here were calculated assuming metallicities
$Z=2\times 10^{-4}$ and $Z=6\times 10^{-4}$. The mixtures are assumed to 
be alpha-enhanced, with $[\alpha/Fe]=+0.4$, so these two chemical 
compositions  correspond, respectively, to [Fe/H]=--2.3 and 
[Fe/H]=--1.83, thus encompassing the chemistry of the most metal poor 
GCs. We have also computed a set of models with metallicity $Z=10^{-4}$, 
to allow a direct comparison with the metal poor models by Karakas \& 
Lattanzio (2007) and Herwig (2004).

The main physical properties of the models for the various metallicities 
are reported in Table \ref{phys}. We also show the results by
Ventura \& D'Antona (2008a) for $Z=10^{-3}$ and Ventura \& D'Antona (2008b)
for $Z=4\times 10^{-3}$. The columns in the table indicate the initial
mass of the model, the duration of the two core nuclear burning phases,
the core mass at the beginning of the AGB phase (when the hydrogen shell
is extinguished after the exhaustion of the central helium), the maximum
luminosities and the maximum temperatures at the bottom of the convective 
envelope reached during the AGB phase, the number of thermal pulses 
experienced by the star before all the envelope is lost, and the maximum
value of the third dredge-up (hereinafter TDU) parameter $\lambda$,
defined as the ratio between the mass dredged up 
in the after-pulse phase and the increase of the
core mass (due to the outwards advancing of the CNO burning shell)
from the previous pulse.

\begin{figure}
\includegraphics[width=8cm]{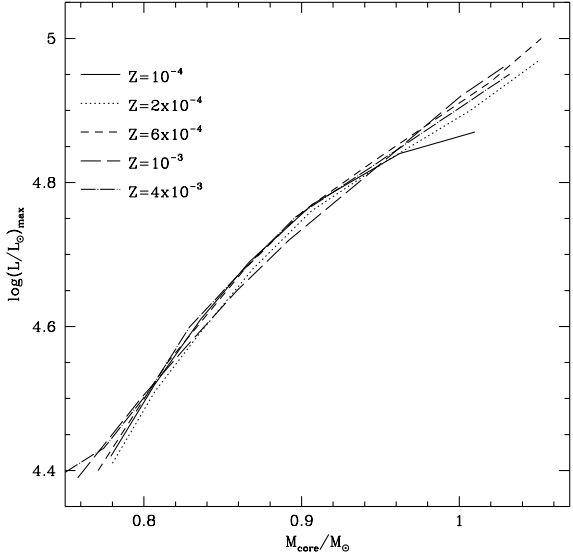}
\caption{Variation of the maximum luminosity attained by the AGB models
as a function of the core mass at the beginning of the AGB phase, when
the CNO shell is extinguished after the core He-burning phase}
      \label{mclum}%
\end{figure}

For a given initial mass M, we see from the 4th column of Table \ref{phys} that
the core mass increases with decreasing Z, a well known result of the stellar
evolution theories; we note a very small difference between the $Z=10^{-4}$
and $Z=2\times 10^{-4}$ sets of models.

Fig.\ref{mclum} shows that, independently of the metallicity, during the
AGB phase the models follow approximately the same relationship between
the core mass at the beginning of the AGB phase ($M_C$) and the maximum
luminosity reached ($L_{\rm max}$) in almost all the range of $M_C$s
spanned by the models; fig.\ref{mclum} shows indeed
a flattening of the $M_C-L_{\rm max}$ trend at the lowest metallicities, 
whereas the higher Z models follow a steeper behaviour. This result can be 
understood on the basis of the variation with Z of the mass fraction of 
the chemical species involved in the CNO cycle, which becomes particularly
relevant whenever strong HBB conditions are reached, for large M: 
the ``extra-luminosity'' gained by the star as a 
consequence of the proximity of the bottom of the convective envelope 
to the CNO shell peak (Ventura \& D'Antona 2005a) grows with the mass
fractions of the CNO elements, hence with Z. This is also the reason
why, for the masses very close to the limit for carbon ignition 
($\sim 6$\msun in the present investigation), the common behaviour that
low Z models attain larger luminosities for a given initial mass is reversed,
as we can see from columns 1 and 3 of Table \ref{phys}; the 6\msun model
of metallicity $Z=10^{-3}$ reaches a higher luminosity than its
lower Z counterparts, despite having a smaller core mass.

\begin{figure}
\includegraphics[width=8cm]{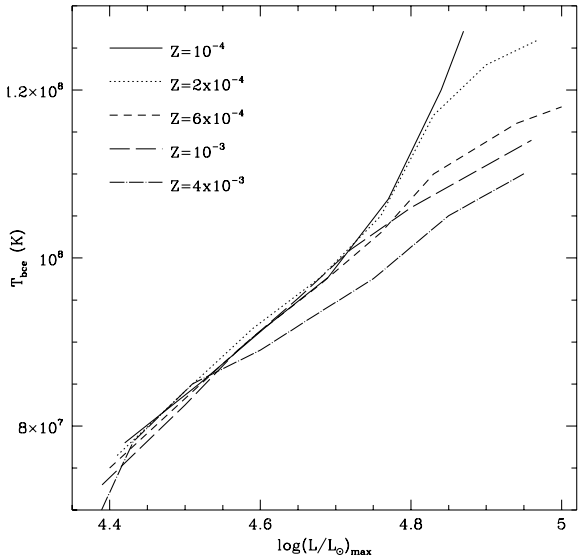}
\caption{The variation of the maximum temperature attained at the
bottom of the convective envelope of the AGB models of various
metallicities as a function of the maximum luminosity reached}
      \label{lumtb}%
\end{figure}

The main difference among the various sets of models presented
here is the temperature at the bottom of the convective envelope:
fig.\ref{lumtb} shows that for a given luminosity the low Z models are
hotter. This behaviour is again a consequence of the smaller
mass fractions of the CNO elements for lower Z, that requires a higher
temperature in the shell to reach the same luminosity.

\begin{figure}
\includegraphics[width=8cm]{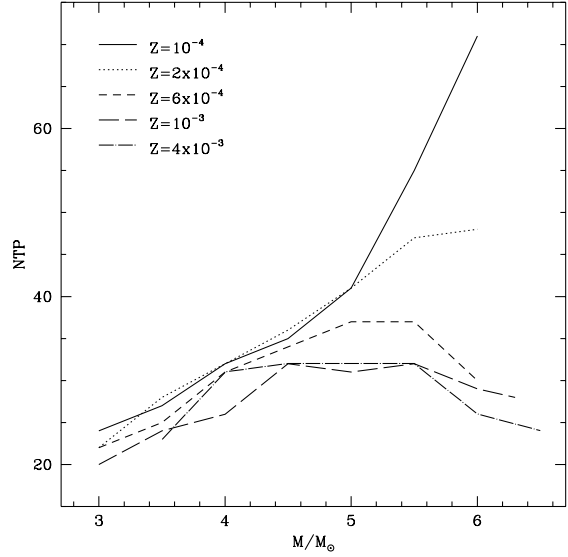}
\caption{Variation with the initial mass of the total number of 
thermal pulses experienced by AGB models before the envelope is
lost}
      \label{mtp}%
\end{figure}

We conclude this analysis discussing the
number of thermal pulses (NTP) experienced by the various masses, shown 
as a function of the initial mass in fig.\ref{mtp}.
For a given mass M, NTP diminishes with Z, as it should be expected 
because higher Z models have larger radii, thus suffer a stronger mass 
loss, that reduces the number of TPs.
The main difference among models with different Z is the general trend of 
the M-NTP relationship: in the lower Z models NTP grows with mass in the
whole range of masses investigated, whereas in the intermediate
Z case NTP reaches
a maximum around 4.5-5\msun and declines with increasing M, because 
of the larger luminosities reached by the higher Z
models for the largest masses, that determines an increase of the mass
loss rate.

We therefore find an important difference between the evolutionary
properties of high mass models of low and intermediate metallicity. 
Low Z models are expected to undergo a more advanced nucleosynthesis at the
bottom of their outer convective zone, and to experience more TPs.

%
\begin{table*}
\caption{Evolutionary properties of intermediate-mass models}             
\label{phys}      
\centering          
\begin{tabular}{c c c c c c c c}     
\hline\hline       
$M/M_{\odot}$ & $\tau_{\rm H}/10^6$ & 
$\tau_{\rm He}/10^6$ & $M_{\rm core}/M_{\odot}$ & 
$\log(L/L_{\odot})_{\rm max}$ & $T^{\rm bce}_{\rm max}$ &  
$N_{\rm pulse}$(NTP) & $\lambda$  \\ 
\hline            
 & & & $Z=10^{-4}$ & & & &     \\
\hline    
   3.0 & 240 & 45.4  & 0.78 & 4.42 & 78  & 24  & 0.7  \\  
   3.5 & 174 & 29.0  & 0.81 & 4.52 & 85  & 27  & 0.7  \\
   4.0 & 132 & 20.3  & 0.84 & 4.61 & 92  & 32  & 0.7  \\
   4.5 & 104 & 14.7  & 0.87 & 4.69 & 98  & 35  & 0.6  \\
   5.0 & 85  & 11.3  & 0.91 & 4.77 & 107 & 41 & 0.5  \\
   5.5 & 70  & 9.0   & 0.96 & 4.84 & 120 & 55 & 0.5  \\
   6.0 & 60  & 7.2   & 1.01 & 4.87 & 127 & 71 & 0.3  \\
\hline        
& & & $Z=2\times 10^{-4}$ & & & &     \\    
\hline   
   3.0 & 248 & 46.5  & 0.78 & 4.41 & 77  & 22  & 0.7  \\  
   3.5 & 179 & 29.8  & 0.81 & 4.51 & 85  & 28  & 0.7  \\
   4.0 & 135 & 20.7  & 0.84 & 4.59 & 92  & 32  & 0.7  \\
   4.5 & 106 & 15.1  & 0.87 & 4.68 & 98  & 36  & 0.6  \\
   5.0 & 86  & 11.4  & 0.91 & 4.76 & 105 & 41  & 0.5  \\
   5.5 & 72  & 9.1   & 0.95 & 4.83 & 117 & 47  & 0.5  \\
   6.0 & 61  & 7.4   & 1.00 & 4.90 & 123 & 48  & 0.3  \\
   6.3 & 55  & 6.4   & 1.05 & 4.97 & 126 & 26  & 0.0  \\
\hline
& & & $Z=6\times 10^{-4}$ & & & &     \\    
\hline   
   3.0 & 265 & 50.3  & 0.77 & 4.40 & 75  & 22  & 0.7  \\  
   3.5 & 188 & 32.2  & 0.81 & 4.55 & 84  & 25  & 0.7  \\
   4.0 & 141 & 22.5  & 0.83 & 4.59 & 91  & 31  & 0.7  \\
   4.5 & 110 & 16.3  & 0.86 & 4.68 & 97  & 34  & 0.7  \\
   5.0 & 89  & 12.4  & 0.90 & 4.76 & 103 & 37  & 0.6  \\
   5.5 & 74  & 9.7   & 0.94 & 4.83 & 110 & 37  & 0.6  \\
   6.0 & 62  & 7.9   & 1.02 & 4.94 & 116 & 30  & 0.3  \\
   6.4 & 56  & 6.6   & 1.05 & 5.00 & 118 & 29  & 0.0  \\
\hline
& & & $Z=10^{-3}$ & & & &     \\    
\hline   
   3.0 & 277 & 55.0  & 0.76 & 4.39 & 73  & 20  & 0.7  \\  
   3.5 & 195 & 34.0  & 0.80 & 4.50 & 83  & 24  & 0.7  \\
   4.0 & 146 & 23.5  & 0.83 & 4.57 & 89  & 26  & 0.7  \\
   4.5 & 113 & 17.3  & 0.86 & 4.65 & 95  & 32  & 0.6  \\
   5.0 & 91  & 12.8  & 0.89 & 4.72 & 101 & 31  & 0.5  \\
   5.5 & 75  & 10.1  & 0.94 & 4.80 & 106 & 32  & 0.5  \\
   6.0 & 63  & 8.2   & 1.00 & 4.92 & 112 & 29  & 0.3  \\
   6.3 & 58  & 7.2   & 1.03 & 4.96 & 114 & 28  & 0.3  \\
\hline
& & & $Z=4\times 10^{-3}$ & & & &     \\    
\hline
   3.0 & 319 & 69.5  & 0.65 & 4.26 & 45  & 25  & 0.7  \\  
   3.5 & 220 & 42.0  & 0.77 & 4.43 & 78  & 23  & 0.7  \\
   4.0 & 160 & 27.8  & 0.80 & 4.51 & 85  & 31  & 0.7  \\
   4.5 & 122 & 19.6  & 0.83 & 4.60 & 89  & 32  & 0.6  \\
   5.0 & 97  & 14.8  & 0.86 & 4.67 & 93  & 32  & 0.5  \\
   5.5 & 79  & 11.4  & 0.89 & 4.75 & 98  & 32  & 0.5  \\
   6.0 & 66  & 9.5   & 0.96 & 4.85 & 105 & 26  & 0.3  \\
   6.5 & 56  & 7.8   & 1.03 & 4.95 & 110 & 24  & 0.3  \\
\hline

\end{tabular}
\end{table*}

\section{The chemical yields}
The chemistry of the ejecta of the AGBs is essential to understand
the role that these stars may play in the context of the pollution
of the interstellar medium; a comparison between the individual
abundances of the mass ejected during their evolution and the 
chemical composition of the stars in GCs with the anomalous chemistry 
provides a good test of the reliability of the self-enrichment scenario. 
Columns 3 to 8 of Table \ref{chem} contain the average abundance 
ratios of the elements mostly investigated in the spectroscopic surveys of 
GCs, in terms of the quantity [X/Fe]=$\log({\rm X/Fe})-\log({\rm X/Fe})_{\odot}$.
The isotopic magnetic ratios of the ejecta are indicated in columns 11 and
12. R(CNO) (shown in col.9) represents the ratio between the global 
C+N+O abundance of the ejecta and the initial value at the beginning
of the evolution, whereas C/O (col.10) indicates the ratio between the
carbon and the oxygen abundance.
The table also includes the results for higher Z models presented in
Ventura \& D'Antona (2008a;b).

\subsection{The CNO elements}
We start our analysis by examining the abundances of the CNO
elements, that during the AGB evolution are modified at the surface
of the stars by TDU and HBB.

\begin{figure*}
\centering{
\includegraphics[width=8cm]{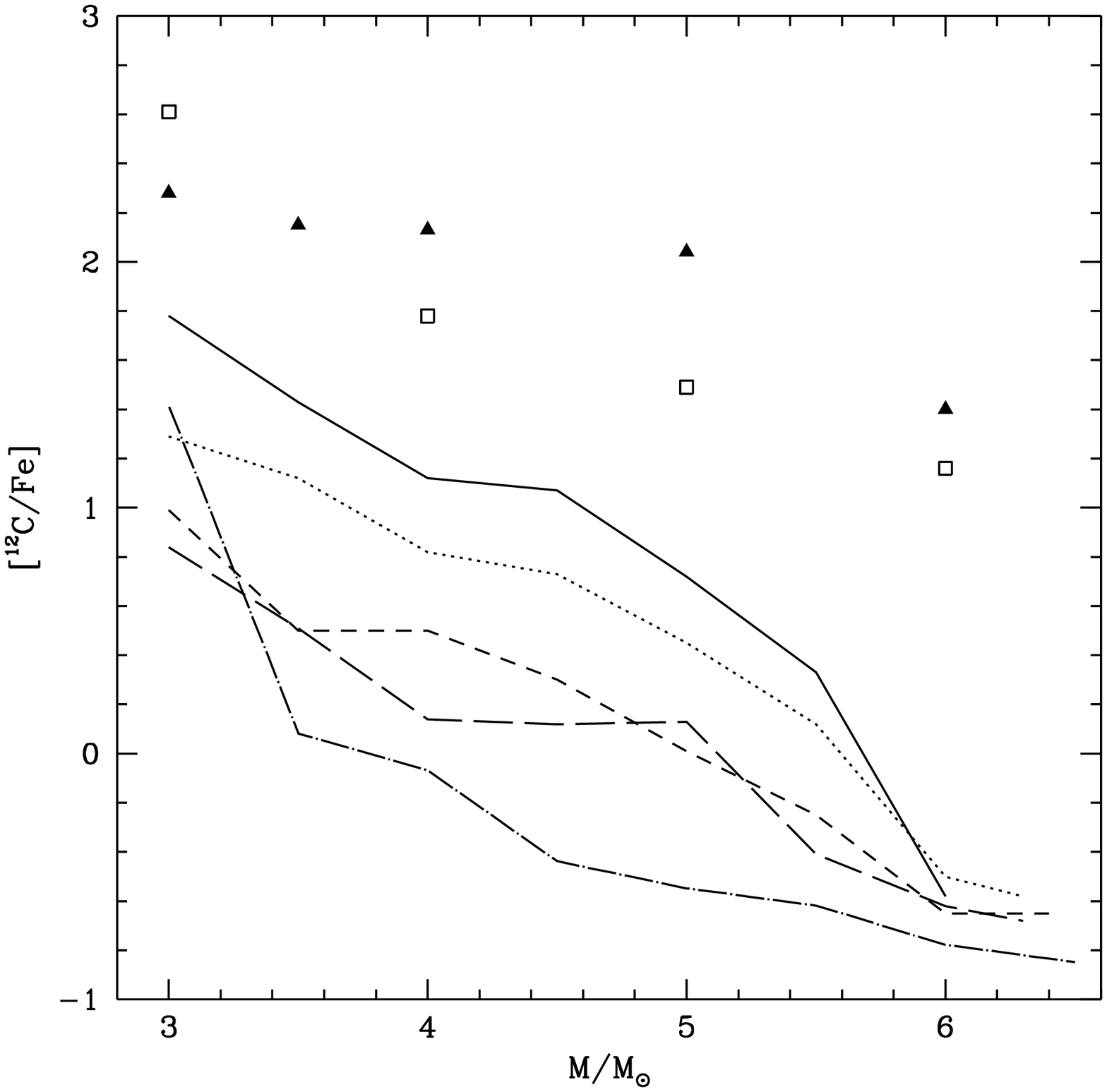}
\includegraphics[width=8cm]{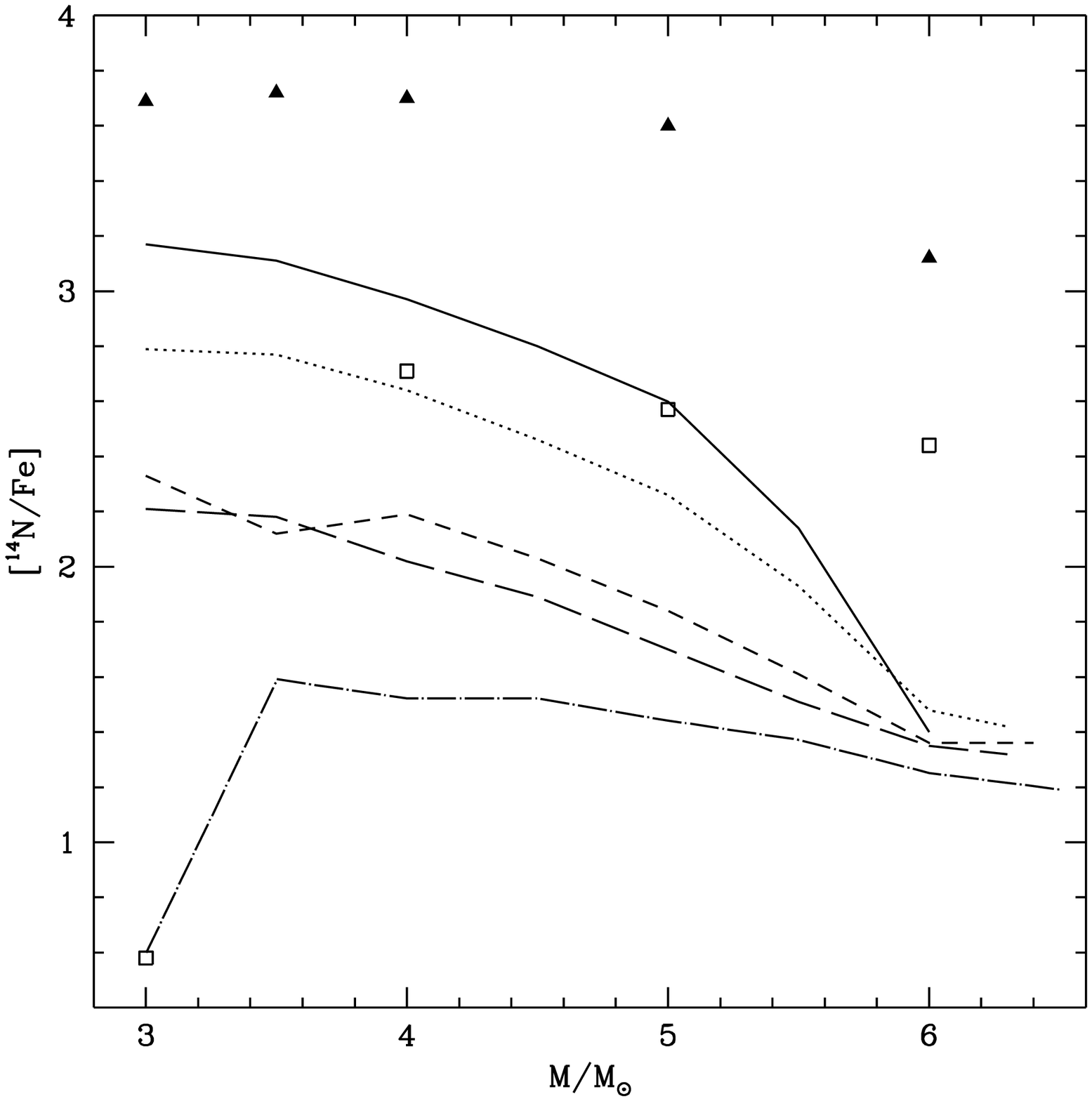}
}
\caption{Variation with the initial stellar mass of the carbon
yield (left, expressed as \cfe) and of the nitrogen yield (right) of 
AGB models of metallicity $Z=10^{-4}$ (solid), $Z=2\times 10^{-4}$ 
(dotted), $Z=6\times 10^{-4}$ (dashed), $Z=10^{-3}$ (long dashed), 
$Z=4\times 10^{-3}$ (dotted-dashed). The full triangles indicate the 
yields by KL07, whereas the open squares refer to the results by
Herwig (2004)}
      \label{cn}%
\end{figure*}

\begin{table*}
\caption{Chemical yields of intermediate-mass models}             
\label{chem}      
\centering          
\begin{tabular}{c c c c c c c c c c c c}     
\hline\hline       
$M/M_{\odot}$ & Y & \cfe & \nfe & \ofe & \nafe & \mgfe & \alfe & R(CNO) & C/O  & $^{25}Mg/^{24}Mg$ & $^{26}Mg/^{24}Mg$\\
\hline            
 & & & & & $Z=10^{-4}$ & & & & &     \\
\hline    
   3.0 & 0.259 &  1.78 & 3.17 &  1.54 &  2.24 &  1.03 & 1.46 & 72.68 &  0.65 &  2.33 & 0.80 \\
   3.5 & 0.281 &  1.43 & 3.11 &  1.32 &  2.19 &  0.90 & 1.30 & 57.03 &  0.48 &  1.65 & 0.58 \\
   4.0 & 0.305 &  1.12 & 2.97 &  1.08 &  1.90 &  0.71 & 1.16 & 39.23 &  0.41 &  1.87 & 0.50 \\ 
   4.5 & 0.321 &  1.07 & 2.80 &  0.83 &  1.51 &  0.54 & 1.19 & 26.77 &  0.65 &  7.23 & 1.35 \\
   5.0 & 0.333 &  0.72 & 2.60 &  0.45 &  0.94 &  0.24 & 1.26 & 14.68 &  0.74 & 16.90 & 2.39 \\
   5.5 & 0.343 &  0.33 & 2.14 & -0.07 &  0.28 & -0.45 & 0.48 &  5.49 &  0.93 &  9.94 & 0.74 \\
   6.0 & 0.351 & -0.58 & 1.40 & -1.43 & -0.20 & -0.53 & 0.20 &  0.95 &  2.62 & 17.02 & 0.42 \\
\hline        
& & & & & $Z=2\times 10^{-4}$ & & & & &     \\    
\hline   
   3.0 & 0.258 &  1.29 & 2.79 &  1.04 &  1.93 &  0.71 &  0.88 & 28.04 &  0.66 &  0.71 & 0.30 \\
   3.5 & 0.279 &  1.12 & 2.77 &  0.96 &  1.88 &  0.72 &  0.99 & 26.07 &  0.54 &  0.85 & 0.31 \\
   4.0 & 0.300 &  0.82 & 2.64 &  0.77 &  1.61 &  0.58 &  0.93 & 18.74 &  0.47 &  1.24 & 0.30 \\
   4.5 & 0.320 &  0.73 & 2.46 &  0.51 &  1.17 &  0.43 &  1.10 & 12.36 &  0.65 &  6.40 & 1.00 \\
   5.0 & 0.315 &  0.45 & 2.26 &  0.19 &  0.70 &  0.20 &  1.23 &  7.51 &  0.68 & 14.47 & 1.92 \\
   5.5 & 0.343 &  0.12 & 1.93 & -0.30 &  0.22 & -0.18 &  0.94 &  3.43 &  0.89 & 16.57 & 0.91 \\
   6.0 & 0.350 & -0.50 & 1.48 & -1.02 & -0.10 & -0.21 &  0.61 &  1.16 &  0.99 & 27.95 & 0.72 \\
   6.3 & 0.354 & -0.60 & 1.42 & -1.05 & -0.14 & -0.02 &  0.69 &  1.00 &  0.97 & 48.11 & 1.02 \\
\hline
& & & & & $Z=6\times 10^{-4}$ & & & & &     \\    
\hline   
   3.0 & 0.253 &  0.99 & 2.33 &  0.80 & 1.51 & 0.59 & 0.63 & 11.10 & 0.45 &  0.37 & 0.18 \\  
   3.5 & 0.263 &  0.50 & 2.12 &  0.52 & 1.43 & 0.50 & 0.54 &  8.29 & 0.35 &  0.43 & 0.13 \\
   4.0 & 0.295 &  0.50 & 2.19 &  0.48 & 1.32 & 0.51 & 0.69 &  7.00 & 0.38 &  0.70 & 0.17 \\
   4.5 & 0.313 &  0.30 & 2.03 &  0.20 & 0.98 & 0.41 & 0.96 &  4.63 & 0.69 &  3.58 & 0.49 \\
   5.0 & 0.329 &  0.01 & 1.84 & -0.12 & 0.58 & 0.26 & 1.16 &  2.89 & 0.49 & 11.79 & 1.51 \\
   5.5 & 0.340 & -0.25 & 1.61 & -0.46 & 0.29 & 0.15 & 1.16 &  1.67 & 0.59 & 20.43 & 1.57 \\
   6.0 & 0.360 & -0.65 & 1.36 & -0.58 & 0.15 & 0.20 & 1.02 &  0.94 & 0.31 & 36.20 & 1.36 \\
   6.4 & 0.360 & -0.65 & 1.36 & -0.53 & 0.12 & 0.21 & 0.97 &  0.95 & 0.27 & 40.40 & 1.32 \\
\hline
& & & & & $Z=10^{-3}$ & & & & &     \\    
\hline   
   3.0 & 0.248 &  0.84  & 2.21 &  0.92 & 1.16 & 0.57 & 0.65 & 9.6  & 0.30 &  0.32 & 0.16  \\  
   3.5 & 0.265 &  0.51  & 2.18 &  0.77 & 1.30 & 0.55 & 0.66 & 7.9  & 0.20 &  0.30 & 0.14  \\
   4.0 & 0.281 &  0.14  & 2.02 &  0.44 & 1.18 & 0.48 & 0.55 & 4.9  & 0.18 &  0.42 & 0.13  \\
   4.5 & 0.310 &  0.12  & 1.89 &  0.19 & 0.97 & 0.43 & 0.85 & 3.1  & 0.30 &  2.19 & 0.31  \\
   5.0 & 0.324 &  0.13  & 1.70 & -0.06 & 0.60 & 0.35 & 1.02 & 2.1  & 0.56 &  8.37 & 0.99  \\
   5.5 & 0.334 & -0.41  & 1.51 & -0.35 & 0.37 & 0.28 & 1.10 & 1.3  & 0.32 & 15.68 & 1.48  \\
   6.0 & 0.343 & -0.62  & 1.35 & -0.40 & 0.31 & 0.27 & 1.04 & 0.97 & 0.22 & 27.87 & 1.56  \\
   6.3 & 0.348 & -0.68  & 1.33 & -0.37 & 0.30 & 0.30 & 0.99 & 0.94 & 0.18 & 30.63 & 1.39  \\
\hline
& & & & & $Z=4\times 10^{-3}$ & & & & &     \\    
\hline
   3.0 & 0.277 &  1.41  & 0.60 &  0.62 & 0.44 & 0.59 & 0.68 & 4.7  & 2.23  & 0.36 & 0.18 \\  
   3.5 & 0.269 &  0.08  & 1.59 &  0.46 & 1.07 & 0.50 & 0.33 & 2.5  & 0.15  & 0.14 & 0.11 \\
   4.0 & 0.281 & -0.07  & 1.52 &  0.30 & 1.17 & 0.48 & 0.32 & 2.0  & 0.15  & 0.17 & 0.10 \\
   4.5 & 0.298 & -0.44  & 1.52 &  0.21 & 1.00 & 0.47 & 0.43 & 1.8  & 0.08  & 0.37 & 0.11 \\
   5.0 & 0.313 & -0.55  & 1.44 &  0.09 & 0.89 & 0.45 & 0.57 & 1.4  & 0.08  & 0.92 & 0.16 \\
   5.5 & 0.328 & -0.62  & 1.37 &  0.01 & 0.76 & 0.43 & 0.70 & 1.2  & 0.07  & 2.22 & 0.28 \\
   6.0 & 0.329 & -0.78  & 1.25 &  0.01 & 0.63 & 0.42 & 0.71 & 1.0  & 0.05  & 4.56 & 0.48 \\
   6.5 & 0.330 & -0.85  & 1.19 &  0.05 & 0.60 & 0.43 & 0.66 & 0.96 & 0.05  & 5.80 & 0.52 \\
\hline

\end{tabular}
\end{table*}

The two panels of fig.\ref{cn} show \c12 and \n14 in the ejecta of 
our models, as a function of the initial 
mass M of the star, for the 5 metallicities discussed. We also report the 
values obtained by Herwig (2004) (hereinfter H04) and by 
Karakas \& Lattanzio (2007) to allow a comparison with sets of models 
provided by different groups: these models have a metallicity $Z=10^{-4}$. 

\cfe\quad is large for low masses, because of the increase of the 
carbon abundance due to the occurrence of the TDU.
The trend with mass is negative, because the smaller is the mass of the envelope,
the larger is the increase of the carbon mass fraction as a consequence
of the TDU. More massive models achieve more easily HBB 
conditions, with the consequent depletion of the envelope \c12 due 
to proton captures at the bottom of the external convective zone;
for masses close to the limit for carbon ignition \cfe\quad is
negative (see col.8 of Table \ref{phys}).

When comparing in fig.\ref{cn} the lines corresponding to different
metallicities, we see that \cfe\quad increases as Z diminishes: this is 
a consequence of the fact that the same quantity of carbon 
dredged-up to the surface determines a larger increase of the carbon 
mass fraction in the envelope of the low Z models. The number of TPs 
and the efficiency of the TDUs are not relevant in determining 
the differences observed
(see Table \ref{phys}). The differences among the various metallicities
vanish at the largest masses, where all sets of models tend
to a limit value of \cfe$\sim -0.7$, independently of Z. In these stars TDU 
is practically not operating, and strong HBB favours carbon destruction, for 
all Z's; the stronger depletion of carbon that is found at low Z's at
the beginning of the AGB phase (due to the higher $\tbce$'s) is 
counterbalanced by the larger carbon equilibrium 
abundances expected when the full CNO cycle is activated.

\begin{figure}
\includegraphics[width=8cm]{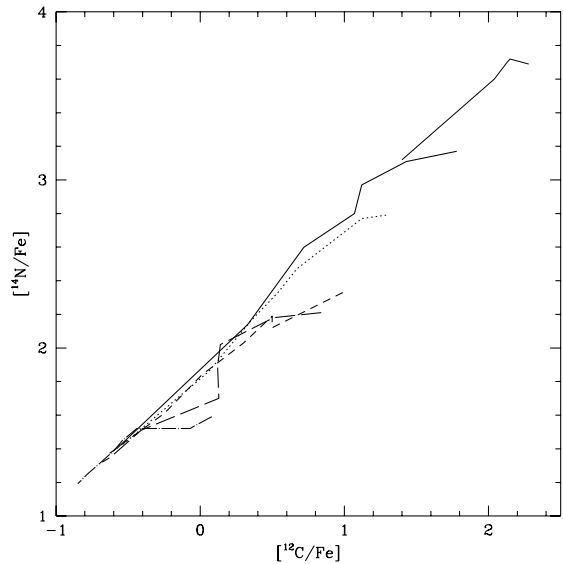}
\caption{The nitrogen content of the ejecta of the AGB models
as a function of the carbon content. The thin solid line 
indicates the results by KL07}
      \label{cvsn}%
\end{figure}

\nfe\quad shows a trend with mass that is similar to \cfe\quad
(see the right panel of fig.\ref{cn}), because \n14\quad
is produced at the bottom of the envelope via proton capture by
\c12 nuclei; we recall that nitrogen production requires only
mild HBB conditions, with no necessity of activating the full
CNO cycle. Contrary to carbon, \nfe~is positive in all cases,
because it is produced and never destroyed. The nitrogen yields
become Z-independent in the range $M\geq 5$\msun, for the same
reasons discussed in the analysis of the carbon yields.

The opposite effects of the HBB for carbon and for nitrogen can be 
detected in the different negative slopes of the variations with mass
of \cfe~and \nfe: the slope of the former is much higher, whereas
the decrease of \nfe~with mass is more modest, as the stronger HBB
acts to destroy carbon and increase \nfe.

\begin{figure*}
\centering
\includegraphics[width=8cm]{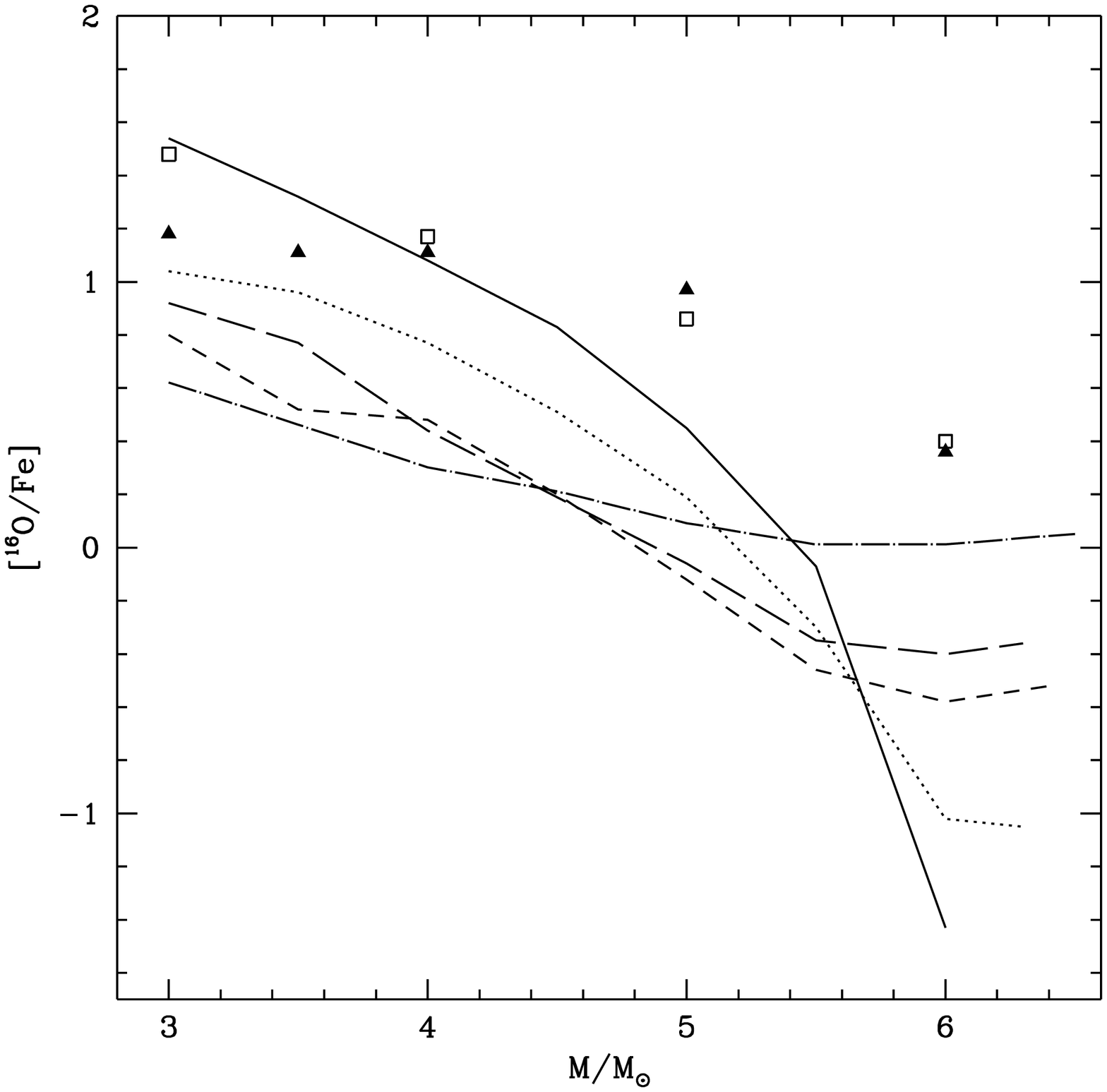}
\includegraphics[width=8cm]{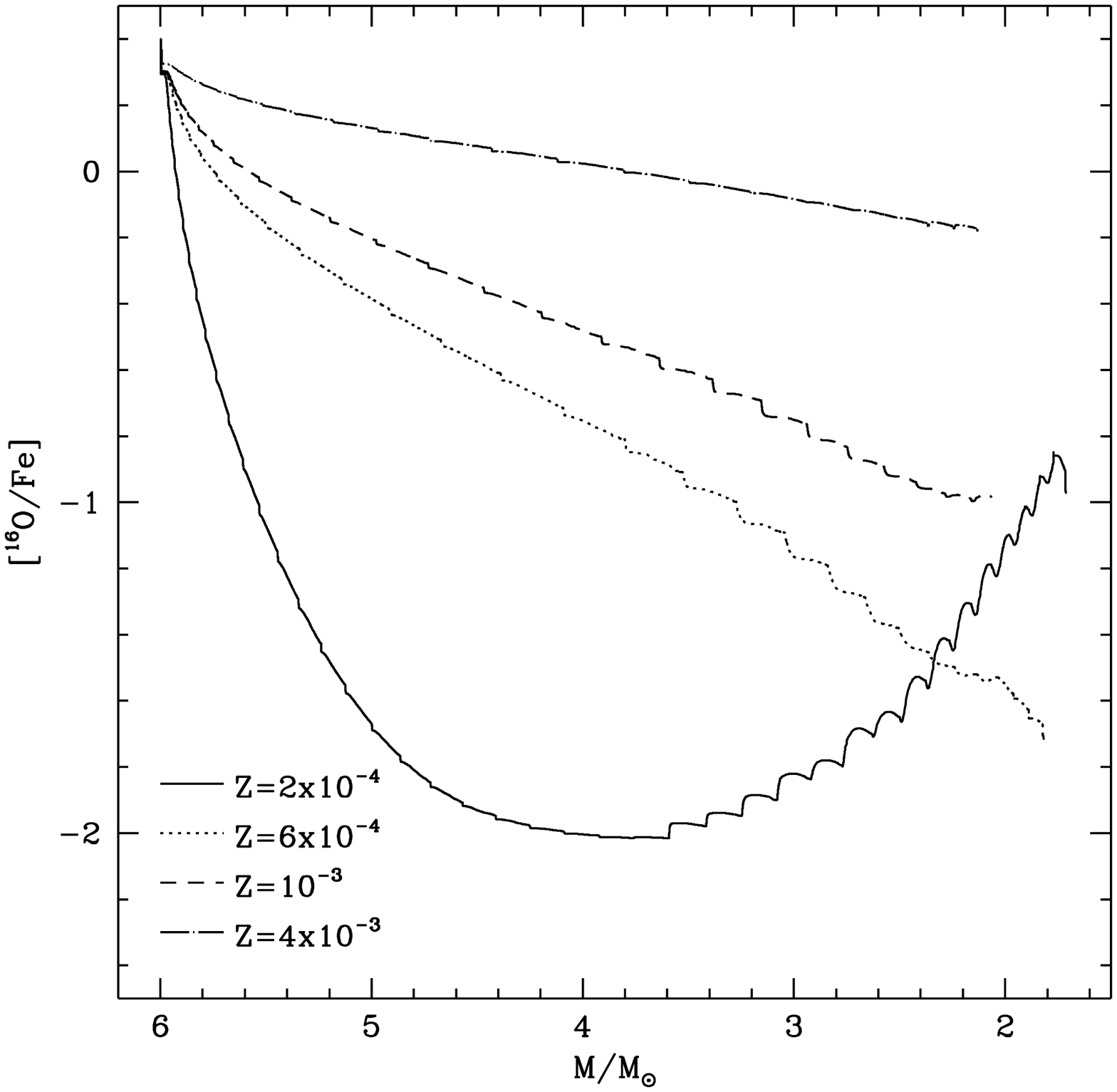}
\caption{Left: the oxygen content of the ejecta of the models discussed
in the paper as a function of the initial mass. The individual
metallicities and the results by H04 and KL07 are indicated 
by different labels, whose meaning is the same as fig.\ref{cn}. Right: 
the variation of the surface oxygen abundance during the evolution of 
massive AGB models of mass 6\msun for various metallicities}
      \label{oxy}%
\end{figure*}

The interpretation of the average oxygen abundance ratios of 
the ejecta, shown in the left panel of fig.\ref{oxy}, and of their trend with
mass and metallicity, is less straightforward than for carbon and 
nitrogen. Oxygen is dredged-up in the phases following the TPs 
only when the TDU is very efficient; HBB leads to oxygen depletion,
but, unlike carbon, this holds only when the full CNO cycle is
activated at the bottom of the convective envelope, when the
temperature exceeds $\sim 70-80$MK. Both factors explain while in
all cases \ofe~decreases with increasing mass, and becomes negative for the
most massive models. 
For the lowest masses of our samples, for which TDU is the main cause
of the change of the surface oxygen, \ofe~increases as Z decreases,
for the same reasons outlined above. On the other hand, when M
increases, the higher temperatures at the bottom of the envelope
attained by the low Z massive models leads to a stronger depletion
of the surface oxygen; this is enhanced by their lower mass loss
rates, because, unlike their higher Z counterparts, they start loosing
most of the mass of their envelope when the surface oxygen abundance
has already considerably diminished (see the right panel of fig.\ref{oxy}). 
The behaviour of \ofe~is thus not monotonic with Z.

In this case, at odds with what found for carbon and nitrogen, the
oxygen yields varies with Z even for large masses, because in the
low metallicity, massive models we expect a very strong depletion of
the surface oxygen, so that the overall reduction factor is almost a 
factor $\sim 10$ larger in the $Z=10^{-4}$ models compared to the 
$Z=10^{-3}$ case.

The sum of C+N+O abundances of the ejecta, shown in the 9th column of 
Table \ref{chem}, diminishes with mass, as a consequence of the 
smaller impact of the TDU for the highest masses, and approaches unity for 
the most massive models; this behaviour is independent of Z. \\
The C/O ratio, reported in the 10th column of Table \ref{chem}, is more
sensitive to Z, and tends to decrease with metallicity. For small masses,
HBB is negligible, and the stronger impact of the TDU in the low Z models
favours a higher C/O ratio (see both panels of fig.\ref{cn}); for the masses 
close to the limit for carbon ignition, the very strong depletion of oxygen 
found in the low metallicity models (see the left panel of fig.\ref{oxy}) 
leads to C/O ratios sligthly below unity 
for $Z\geq 2\times 10^{-4}$, and $C/O>1$ at $Z=10^{-4}$. \quad
$Z\sim 2\times 10^{-4}$ \quad is the threshold metallicity below which 
O-poor yields, leading to the situation where C/O exceeds unity at the
surface, are to be expected.

We conclude this discussion with a comparison of our yields with 
those by KL07 and H04. 
We see from fig.\ref{cn} and \ref{oxy} that our CNO abundances are 
sistematically lower than those by KL07, so we find a much greater 
CNO enhancement when the KL07 models are adopted; also, \cfe~is always 
positive in the KL07 case. The key role in this context is played by the
treatment of convection: in the models presented here convection was 
modelled according to the FST treatment, whereas the KL07 models were 
calculated by means of the MLT description, that provides a much less 
efficient description of the  convective instability. As discussed in 
Ventura \& D'Antona (2005a), a more efficient convection model favours 
larger temperatures at the bottom of the convective envelope, larger 
luminosities, shorter life-times, and a faster loss of the mass of the 
envelope. The above explanation likely leads to the situation
observed, where the KL07 models experience more TPs than our models.
For example, our 3, 4, and 5\msun $Z=10^{-4}$ models experience, 
respectively, 24,32,41 TPs, compared
to the 40,76 and 138 TPs suffered by the same masses by KL07.
The TDU episodes following these TPs results in large amounts 
of carbon being dredged-up to the surface, with the consequent 
increase of the abundances of all the CNO elements. The difference 
between our yields and those by KL07 increases with mass,  
because the higher temperatures attained at the base of the convective 
envelope lead, for a given mass, to a more efficient HBB.

The C and N yields by H04 are lower than those by KL07, because the 
H04 models experience a much smaller number of TPs due to the higher mass 
loss rate adopted (that is a Bl\"ocker (1995) law, enhanced by a factor 5 
compared to ours), thus reducing the amount of carbon that is dredged-up 
to the surface (and that can be eventually converted to nitrogen). 
Compared to ours, the CN H04 yields are higher, the difference increasing 
with mass, because of the stronger HBB conditions experienced by our models 
at a given mass (for example, the average temperature at the bottom of the 
convective envelope for the 6\msun model is 120MK in our case, and 100MK 
in H04). The difference in the oxygen yields (see the left panel of 
fig.\ref{oxy}) also increases with mass between us and H04 models.
In the interpretation of the differences between our results and those
by H04, it is important to stress that the different efficiency of
the TDU plays also a role: in our models no extra-mixing is assumed from
the bottom of the convective envelope, whereas an exponential overshooting
is adopted in the H04 models.

The differences due to the metallicity and to the
treatment of convection can be more easily understood when the yields
of the models are shown on the C-N plane; this will also be of help
when discussing the self-enrichment scenario on the basis of the
observed C-N abundances of SGB stars in M15. The \nfe~values of the
5 sets of models discussed here and the KL07 models are shown as a
function of \cfe~in fig.\ref{cvsn}. We note that all the models trace
approximately a straight line in the C-N plane, to confirm that CN
cycling operates at the bottom of the envelope. The KL04 models are 
located in the right-upper portion of the plane, whereas our models 
occupy the lower region. In agreement with the discussion following
the presentation of the carbon and nitrogen yields, we find that,
independently of Z, all the curves corresponding to the different
metallicities converge to the same locus on the plane, that is
(\cfe,\nfe)$\sim (-0.7, 1.3)$. This does not hold for the KL07 models,
for which a minimum increase of carbon and nitrogen of, respectively,
a factor of $\sim 20$ and $\sim 1000$ are found.

\subsection{Sodium}
The sodium content of the ejecta of AGBs is an essential result in
the way to understand the star-to-star differences of the GCs stars
in the context of the self-enrichment scenario: the oxygen-sodium
anticorrelation is by far the most investigated anticorrelation
that was confirmed by deep spectrospic analysis performed on many GCs
(Carretta et al. 2006).

\begin{figure}
\includegraphics[width=8cm]{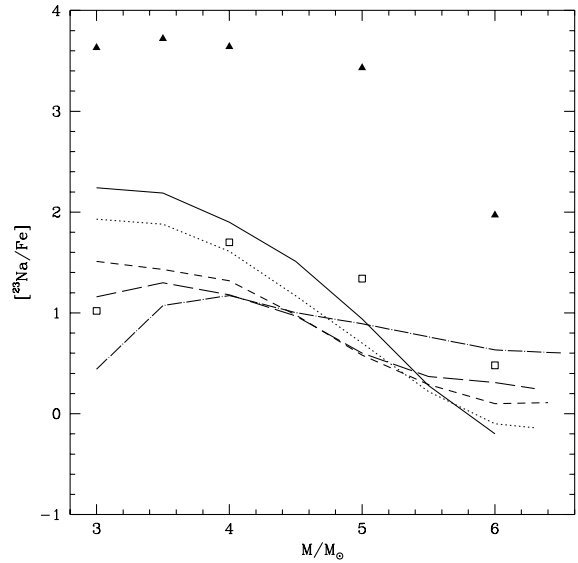}
\caption{The sodium content of the ejecta of the AGB models
as a function of the carbon content. The meaning of the different
labels is the same as in fig.\ref{cn}.}
      \label{sodio}%
\end{figure}

The surface abundance of sodium is determined by the two
processes that change the surface chemistry of AGBs. TDU tends
to increase the surface sodium mass fraction via dredging-up of
\ne22 from the ashes of the $3\alpha$ burning shell; HBB favours
a further increase of the surface sodium as far as the temperature
at the bottom of the envelope does not exceed $\sim 70$MK, above
which the destruction channels dominates.
These considerations allow us to understand the tracks shown in
fig.\ref{sodio}, that show the content of sodium (in terms of \nafe)
as a function of the initial mass of the ejecta of the various sets
of models presented here; we also show the results by KL07  
and H04.

The negative trend of \nafe~vs M found for all Z's
can be explained, as it was for the carbon and oxygen,
by the fact that in the lowest masses models we see mainly the 
effects of TDU, acting to enhance the surface sodium, whereas at
large M's HBB is the dominant mechanism, leading to smaller sodium
contents the larger is the temperature at the bottom of the convective
envelope. As for oxygen, we do not find a monotomic behaviour with
Z, because for low metallicities the bottom of the surface convective
zone of massive AGBs becomes so hot to favour sodium depletion.

The comparison of our results with the yields by KL07 confirms that
it is the temperature in the innermost layers of the outer convective
zone to drive physically and chemically the evolution of AGBs. KL07
sodium yields are sistematically higher than ours, due to the more numerous
TDU episodes, that carries to the surface more \ne22, available to be
converted into sodium during the quiescent CNO burning phase; the
milder efficiency of HBB prevents the strong depletion of the surface
sodium in the more massive models found in our computations.

Our sodium yields are more similar to those by H04; the slope of
the \nafe vs. mass relation is steeper in our case, due to the
strong sodium depletion achieved in our high mass models.

Before proceeding further, it is mandatory to remember that,
unlike the CNO elements, the reliability of the results
obtained for sodium is extremely low, because the cross-sections
of the three key-reactions relevant to determine the sodium yield, i.e.
the creation channel $^{22}$Ne(p,$\gamma$)$^{23}$Na, and the two
destruction channels $^{23}$Na(p,$\gamma$)$^{24}$Mg and 
$^{23}$Na(p,$\alpha$)$^{20}$Ne, are uncertain by up to 3 orders of 
magnitude (Hale et al.2002; 2004).
Ventura \& D'Antona (2008a) showed that according to the cross-sections
adopted, the average sodium mass fraction of the yields of their most
massive intermediate metallicity AGB models would be increased or diminished
compared to the initial abundance (see their fig.9). A more detailed
investigation on this topic was made by Izzard et al. (2007), who
showed that the sodium yields of AGB models of low metallicity was highly
uncertain, and that the poor knowledge of the cross-sections of the relevant
proton capture reactions determine an uncertainty associated to the
expected sodium yield of the order of $\sim 10-100$.

\subsection{Aluminum}
Aluminum is seen to correlate with sodium and to be anticorrelated to
oxygen and magnesium in stars belonging to GCs where clear star-to-star differences
are observed. Among all the light elements involved in the commonly
studied abundance patterns, aluminum is the species showing the largest
spread, the maximum detected abundances being of the order of
\alfe$\sim 1$ in all the stars showing a great depletion of oxygen and
an enhancement of sodium. Gratton et al. (2001), in an
analysis of the surface chemistry of TO and SGB stars in the GC NGC~6752,
found stars with \alfe=1. A similar aluminum enhancement was also 
detected by Sneden et al. (2004) in giants of low and high gravity in
the two GCs M3 and M13. A more recent work by Smith et al. (2005) on M4,
unfortunately limited to giants, evidentiated the presence of aluminum
rich stars, with \alfe=0.8. Finally, Sneden et al. (1997) detected a
few stars in the GCs M15 and M92 (both clusters have a metallicity more 
appropriate to this investigation) with \alfe$\sim 1$, although even in
this case the study is focused on bright giants.

Aluminum is produced in AGBs by HBB via the activation of the Mg-Al chain;
this requires temperatures of the order of 80MK.
TDU also determines an indirect aluminum enhancement, because 
the two magnesium isotopes produced in the 3$\alpha$ burning shell are
convected to the surface, where they synthesize aluminum via proton capture.
Ventura \& D'Antona (2008a) found that, when the upper limits for the
cross-sections of the proton capture reactions by the heavy magnesium 
isotopes are adopted, the most massive among their AGB
models of intermediate metallicity produce great amounts of aluminum,
with an average increase of a factor $\sim 10$, in agreement with the
observations of GCs of that chemistry, like M3 and M13 (see the corresponding
lines in Table \ref{chem}). 
The results by Ventura \& D'Antona (2008b) (obtained by using the
same upper limits for the above mentioned proton capture reactions, as
also in the present investigation) confirmed the possibility of 
producing aluminum at the surface of
massive AGBs also for more metal rich models, though in this case the
maximum enhancement found was \alfe$\sim 0.7$.

The present results confirm the outcome of the above mentioned
investigations, as can be seen by noting the results concerning the
magnesium and aluminum yields in Table \ref{chem}. Here we also
show the $^{25}$Mg/$^{24}$Mg and $^{26}$Mg/$^{24}$Mg ratios, that
are extremely dependent on the assumed rates of the proton capture
reactions by the two heavy magnesium isotopes.
We note in Table \ref{chem} that for low metallicity models with
$Z < 6\times 10^{-4}$ the trend \alfe~vs M is not monotonic
for the whole range of masses investigated: for $Z=2\times 10^{-4}$ 
we find an aluminum enhancement by a factor of $\sim 10$ for masses 
$M\leq 5$\msun, whereas slightly lower values (\alfe $\sim 0.6-0.8$) 
are predicted for the masses close to the limit for carbon ignition. 
This effect is due to the very high temperatures achieved by these
models at the bottom of their external convective zone, that becomes
sufficient to activate efficiently the proton capture reaction by
$^{27}$Al nuclei. On the basis of these computations, the maximum
enhancement reached by the most massive models of low metallicity
is of the order of \alfe$\sim 0.7$.

\section{The observed abundance patterns in low metallicity
Globular Clusters stars}
Ventura \& D'Antona (2008a,b) compared the yields of AGB models
with the observed abundance patterns evidentiated by deep spectroscopic
investigations of GC stars with \feh$\sim -1.3$ and \feh$\sim -0.7$. 
We extend here the analysis to lower Z, making use of the new models 
presented in the previous sections.

Unfortunately, the data for the lowest metallicity clusters are quite scarce, and
only few data are available for non evolved or scarcely evolved stars.

We will use data for the clusters M15 and NGC~6397. Harris (2003) 
lists [Fe/H]=--2.26 for M15, and [Fe/H]=--2.0 for NGC~6397. For the latter cluster, we also have a measure of
the $\alpha$--enhancement, [$\alpha$/Fe]=+0.34$\pm$0.02 (Gratton et al. 2003), while 
Str\"omgren photometry provides [Fe/H]=--1.83 $\pm$0.04 and  
[$\alpha$/Fe]=0.3 (Anthony-Twarog \& Twarog 2000). In our $\alpha$--enhanced models,
Z=2$\times 10^{-4}$\ corresponds to [Fe/H]=--2.3, and Z=6$\times 10^{-4}$\ 
corresponds to  [Fe/H]=--1.82, so we will use these two sets for the respective 
comparisons.

Sneden et al. (1991) detected a clear spread in the surface oxygen 
abundances of giant stars in the two metal-poor GCs M15 and M92. 
Their fig.12 shows that there may be some correlation of the
oxygen abundance with the evolutionary status of 
the individual objects. More luminous stars, closer to the RGB tip, 
show the lowest oxygen abundances, although a wide spread is
found at all luminosities. For the few stars for which also the 
nitrogen abundance was measured a N-O anticorrelation exists, indicating 
the presence of material processed by proton capture nucleosynthesis.
A more complete analysis of M15 giants, focused on the abundances of 
oxygen, sodium, magnesium and aluminum, was presented by Sneden et al. 
(1997). The main result of this investigation is that oxygen is anticorrelated
with sodium, and magnesium anticorrelates with aluminum, indicating again
the signatures of proton capture processing.
Carretta et al. (2005) examine the CNO abundances for several clusters,
including the low Z cluster NGC~6397. Data for dwarf and subgiants are available
for CNO and Na. Finally, Cohen et al. (2005) present an analysis of a large 
sample of spectra of subgiants stars at the base of the RGB of M15, focused 
on the abundances of carbon and nitrogen. 

A straight comparison between the yields of our models and the observations
can be used as a test of the self-enrichment scenario only in the 
case that the observed abundances have not been altered by any in-situ process,
to ensure that the chemistry observed is the same with which the stars formed.
This is surely the case for the Cohen et al. (2005) and Carretta et al. (2005)
data, because the sample of stars observed are in evolutionary stages for which 
advanced nucleosynthesis in their internal regions can be ruled out on the basis 
of their low internal temperatures. 
The same conclusion does not necessarily hold for the two 
surveys by Sneden and coworkers, that involve only bright giants, close to
the RGB tip. Canonical stellar models predict the first dredge-up as the only 
episode up to the helium flash that can alter their surface chemistry, 
changing only the abundances of the two carbon isotopes and of nitrogen, leaving 
unvaried the abundances of heavier nuclei; yet, the investigations by Sweigart \& 
Mengel (1979), and Cavallo et al. (1998) showed that in low Z ($\leq 5\times 10^{-4}$) 
giants rotationally-driven meridional circulation currents, if present, can
penetrate deeply into internal regions, because the entropy barrier associated
with the drop of the hydrogen content, that prevents the inwards penetration
of the surface convetive layer, is more internal. More recently, Eggleton et al. 
(2008), based on 3D numerical simulations, found that a deep mixing mechanism
associated to a small molecular weight inversion must be operative in all
low-mass giants.
An observational indication that oxygen can attain lower abundances in luminous 
giants came from the early M13 data by Sneden et al. (2004). Furthermore, 
Carretta et al. (2006) showed that the lowest oxygen abundances in NGC~2808 are 
present only among giants. Based on these indications,
D'Antona \& Ventura (2007) reproduced the most extreme oxygen and sodium
abundances by  applying deep extra-mixing from the bottom
of the surface convective envelope of giant stars, based on the assumption
that stars born from the ashes of an early generation of massive AGBs should
have a higher helium mass fraction, that would decrease the height of 
the above mentioned entropy barrier that prevents the inwards penetration of 
the convective envelope.
These results indicate that the surface chemistry
of low-Z giants can be altered during their RGB evolution,
particularly after the CNO burning shell crosses the chemical discontinuity
left behind by the first dredge-up episode. Based on this, we proceed to a
full, detailed comparison between our theoretical yields and the observed
abundance patterns mainly for the case of the O--Na anticorrelation in
NGC~6397 and in the case of the M15 C and N data by Cohen et al. (2005).
In spite of the warnings made above, we will also comment on the O, Na, Mg and 
Al data by Sneden et al. (1997) for M15 giants.

\subsection{The O--Na anticorrelation in NGC~6397}

\begin{figure}
\includegraphics[width=8cm]{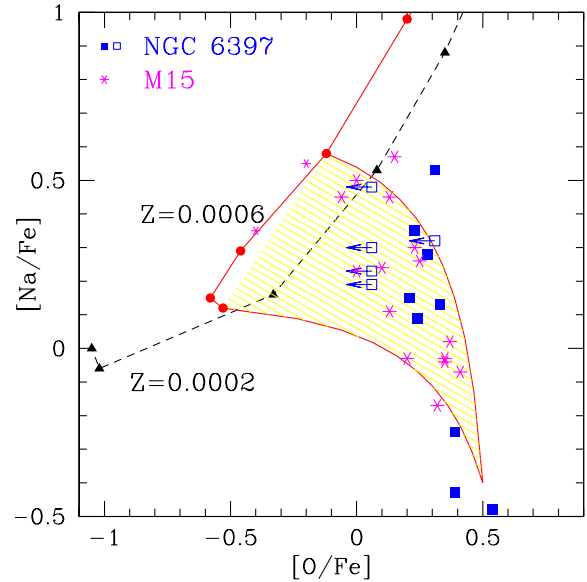}
\caption{We show the Na--O data for TO and SGB
  stars belonging to the GC NGC~6397, according to Carretta et
  al. (2005). Data for M15 giants by Sneden et al. (1997) are also plotted as asterisks.
  The triangles and dashed line are the yields of the Z=0.0002 models, while the dots and
  full line represent the yields of the Z=0.0006 models (6.4, 6.0, 5.5, 5.0
  and 4.5\msun models are plotted in both cases). The latter metallicity is
  the most adequate for NGC~6397, considering the $\alpha$--enhancement. 
  In order to reproduce the data points, it is necessary to consider dilution curves between
  the 6.4 and 5\msun\ yields and an initial composition represented by the vertex of the
  cone, and correspondig to the first generation composition. 
  }
      \label{ona6397}%
\end{figure}

Figure \ref{ona6397} shows the Na vs. O data for NGC~6397 subgiants and dwarf stars
listed by Carretta et al. (2005). Upper limits are indicated by open squares with
arrows. We also report the data of M15 giants by Sneden et al. (1997). An exam of their
Figure 4 shows that also the data for M~92 occupy the same region of the Na--O plane.
The Na--O yields of our models from table 2 are reported in the figure. It is evident that
the models predict much lower oxygen abundance, {\it if the SG stars were
formed from undiluted matter from the AGBs}. Nevertheless we can reproduce the abundances
of these clusters if we make the hypothesis that the matter from the ejecta of
stars of initial mass 6.4 to 5\msun is diluted with pristine matter at the level of
$\sim 50$\%\footnote{The dynamical models by D'Ercole et al. (2008) have shown that the pristine gas remaining in the outskirts of the globular cluster at the end of the supernova epoch, falls back into the cluster core, where it mixes with the AGB ejecta and forms the
SG stars. Depending on the cluster mass and history, the SG can be totally or in part made up
exclusively by the AGB ejecta, or its formation may directly start in the mixed gas.}. The 
two solid curves in the figure represent the composition of matter having the
starting abundance of the 6.4 and 5.0\msun\ ejecta, diluted at different percentages with
pristine matter.
If we make the hypothesis of dilution, the helium content of the SG stars can be obtained by
considering the helium abundance given in Table 2 for the 6.4 and 5\msun\ stars 
(Y=0.33-0.36) and diluting it by half with matter having the Big Bang initial abundance Y=0.24. 
The resulting  Y$\sim$0.28--0.30, is consistent with the ``short" blue HB of this cluster: 
an extremely blue HB would be obtained if the AGB matter forming the
SG had been undiluted. In fact, if the helium content of the SG were  
as high as Y=0.36, both a very blue HB (e.g. D'Antona \& Caloi 2004) and a split of the 
main sequence (Piotto et al. 2007) as in the cluster NGC~2808 would appear in the data.

The few M15 data shown in Fig. 8 refer to giants, and our models for the 
metallicity Z=2$\times 10^{-4}$ appropriate 
for M15, predict SG stars with low sodium and very low oxygen abundances 
in these same stars, depending on reaction rates. There are currently not enough 
data available to test these predictions.
As for the sodium, we have already noticed that the initial value of 
$^{20}$Ne in the models and the cross sections can affect very much the final abundance. 
On the contrary, the oxygen yield can not be changed, unless we re discuss the efficiency 
of convection in our models. In the following Section, discussing the C and N abundances 
in M15, we will see that some of its subgiants are consistent with the abundances in our 
most massive models, so that we should expect to find very low oxygen abundances is these 
same stars, but data are not available to falsify the model.

\subsection{The C--N anticorrelation in M15}

\begin{figure}
\includegraphics[width=8cm]{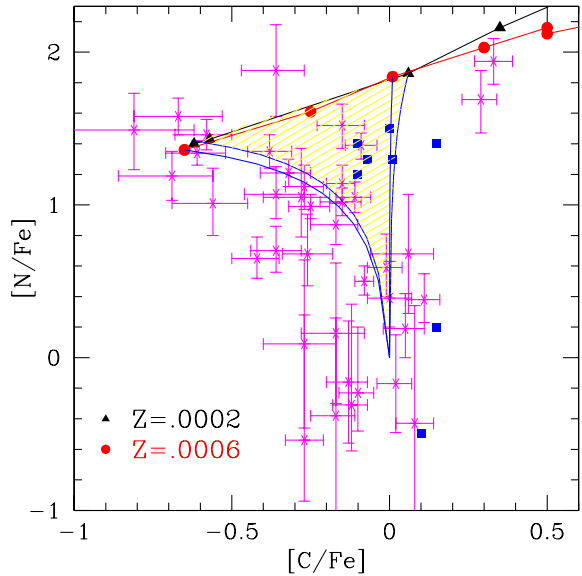}
\caption{Data points for TO and SGB
  stars belonging to M15 (Cohen et
  al. 2005) are shown as asterisks with error bars. 
  Data for NGC~6397 from Carretta et al. (2005) are shown as full squares.
  The lines with triangles ($Z=2\times 10^{-4}$)  and dots  ($Z=6\times 10^{-4}$) 
   are the theoretical abundances, for masses   $\leq 6.4$\msun. 
  The cones corresponds to the regions allowed by dilution of the ejecta from masses
  6.4$\leq M/M_\odot \leq$5.5   with pristine matter having the 
  composition of the cones vertex. }
      \label{cnm15}%
\end{figure}

The main finding by Cohen et al. (2005), shown in fig.\ref{cnm15} and
based on the abundances included
in their Table 2, is the existence of a clear anticorrelation between the surface
abundances of carbon and nitrogen: a group of star (on the left-upper part of
fig.\ref{cnm15}) shows a large enhancement of nitrogen (\nfe$\sim 1.4$) and a
strong depletion of \c12 ((\cfe$\geq -0.8$). 
In Fig. 9 we also add  Carretta et al. (2005) data for the cluster NGC~6397.
Overimposed to the observed points
is our dilution region, obtained assuming different degrees of dilution between 
matter with the chemistry of  models with $Z=2\times 10^{-4}$ and $Z=6\times 10^{-4}$ ,
and masses between 6.4 and 5.5\msun, 
and gas with the original ``standard " chemistry\footnote{The mass range from which we
think the polluting matter comes from is quite small. This is indeed the reason why models for
the formation of multiple populations in globular clusters require that the first generation initial mass
must  be much larger than the mass observed today (e.g. D'Ercole et al. 2008)}.
We see that the theoretical carbon and nitrogen abundances reproduce
satisfactorily the observed patterns. 

The few data for NGC~6397 are well consistent with the dilution expected
from the oxygen and sodium data shown in Fig. 8. On the contrary, 
a direct comparison between Figure 8 and 9 can not be made for M15, as the oxygen and
sodium abundances refer to different stars. Nevertheless, the C and N data
require that the stars with the lowest carbon abundances are made up from undiluted
ejecta of the most massive AGBs. We predict that these stars should have very low oxygen abundances.

In their comparison between their
observed abundances and the previous models by Ventura et al. (2002) (that are in
good agreement with the present yields) Cohen et al. (2005) argue that the agreement
is only qualitative, as the models fail to reproduce the great nitrogen
enhancement evidentiated by the difference between their lowest and highest
values (\nfe$\sim -0.5$ and \nfe$\sim 1.5$ in their fig.4). This conclusion is
actually somewhat misleading. In fact, nitrogen, whatever mixture is used, is the least
abundant among the CNO elements, so that the final nitrogen yield (proportional to the overall C+N+O abundance , under strong HBB conditions) 
turns out to be almost independent of the initial nitrogen adopted, provided
that the same carbon and (expecially) oxygen are used. We confirmed this conclusion
by calculating a 6\msun model with metallicity $Z=2\times 10^{-4}$ with an
initial abundance of nitrogen \nfe=--0.5: we find that the N yield as function of 
the mass is practically the same (0.05 dex lower) as in Table \ref{chem}.

Interestingly, Cohen et al. (2005) compare their C-N trend for M15 with those
from other clusters with larger metallicity, and find that, independently of
\feh, the observations encompass the same range of values of \cfe~and \nfe.
This result is consistent with our finding that for the models experiencing
strong HBB involving full CNO burning and in which the effects of TDU are
negligible the yields of carbon and nitrogen become Z-independent, and converge
to the most extreme values detected by the spectrospic investigations quoted by 
the authors.

\subsection{The Mg -- Al anticorrelation}
Very few data are available, but Sneden et al. (1997) list Mg and Al abundances 
for the same sample of M15 giants for which we have shown the O--Na data
in Fig.8. Figure 3 in their paper shows the trend of Na, Mg and Al with respect to 
[O/Fe] in this sample. Our data in Table 2 for Z=2$\times 10^{-4}$ \ are consistent 
with these trends: low oxygen corresponds to low magnesium and high sodium. Also in 
this case, however, we must NOT consider the most extreme yields of the 6.4 and 6.0\msun, 
as no stars with very low oxygen (and with low sodium, but we 
regard this result less reliable) have yet been found in the cluster. The lack of low oxygen 
stars in M15 requires an explanation, and however stars with very low oxygen should be 
searched for in very low metallicity clusters. If none is found, 
either the present models achieve too strong HBB, or the problem is shifted to the 
modalities of formation of the lowest metallicity clusters, in the sense
that some other mechanism prevents the formation of SG stars directly from
the ejecta of the most massive AGBs.

\subsection{The C/O ratio in low metallicity environments}
Table 2 shows that the C/O ratio in the ejecta of the lowest metallicity
AGB models become increasingly larger as the mass of the model increases, 
and it is definitely larger than one in the 6\msun\ model of Z=10$^{-4}$. 
Thus, during their life, the most massive AGBs of low metallicity become Carbon stars,
although this characteristic is not due to dredge up of carbon, but to the very strong oxygen
depletion due to proton captures on oxygen nuclei during the hot bottom burning. 
The occurrence of Carbon--star stages of evolution in massive, low Z, AGBs has been 
already discussed in Ventura et al. (2002) \footnote{A different mechanism producing 
a C--rich stage in massive AGBs is described by Frost et al. 1998. In that case, the 
occurrence is due to the combined action of the third dredge up and of mass loss that 
reduces the action of HBB in the latest phases of the AGB life, and observational 
counterparts may have been identified in the obscured, C--rich stars of high bolometric 
luminosity (van Loon et al. 1999).}.
A full explanation of the O--Na anticorrelation requires that 
AGB matter is diluted with pristine cluster matter in most of the SG
stars (Ventura \& D'Antona 2008), so it is not clear how large
are the C/O ratios we should expect in the low Z clusters. A full discussion of these results
is postponed to an analysis including results of modeling at smaller metallicities. Here 
we limit it to a few speculative issues.
\begin{itemize}
\item Table 2 shows that the ratio C/O from the massive AGB ejecta becomes larger going
from higher to lower metallicity clusters; this matter is also expected to
be extremelly oxygen--poor. This theoretical prediction requires observational
verification. 
\item  it is curious that clusters of metallicity $< 2 \times 10^{-4}$, in which
we should positively predict that some second generation stars would have C/O$>$1,
do not exist. There are, however, many carbon-rich halo stars with metallicities 
smaller than $10^{-4}$ (e.g. Beers \& Sommer-Larsen 1995, Beers and Christlieb 2005). 
Do carbon rich grains form from the carbon not locked into CO, even if  the 
metallicity is so low? Does this affect the formation of  second generation stars? 
Dust formation in the envelopes of giants is found in C--rich models of stellar 
envelopes at least down to the metallicity of the Small Magellanic Cloud 
(Wachter et al. 2008).
\item in the ejecta having C/O$>$1,  the carbon abundance is small 
([C/Fe]$<$0, Table 2), so that the possible observational counterparts can not 
be looked for among CEMP (carbon enhanced metal poor) stars, defined as stars 
for which [C/Fe]$>$1 (Beers and Christlieb 2005). These models also have 
extremely low C/N ratios, as nitrogen is very large due to HBB (both carbon and 
oxygen contributing to it). These are then NEMPs (nitrogen enhanced metal poor 
stars) in the definition by Johnson et al. (2007), who failed to find any of 
such stars by searching among extremely metal poor stars. Notice that a ratio 
[C/N]$\simeq -2$ is expected from Table 2, but not even less extreme ratios, 
like [C/N]$\simeq -1$ ---as expected from the models by Herwig (2004)--- were 
found in this survey. The presence of enhanced C and N in extremely metal poor 
low mass stars presently observed is generally attributed to mass transfer from 
a previously evolving AGB companion (e.g. Lucatello et al. 2005).  Johnson et 
al. (2007) attribute the lack of NEMPs to the shortage of binaries with the 
extreme mass ratios required to produce them.
\item the presence of large C/O ratios ---and of total low C and O abundances, could
be detected by looking at the properties of gas and dust in the circumnuclear region
of the very high redshift QSO (e.g. Maiolino et al. 2004).
\end{itemize}

\section{Conclusions}
We present new evolutionary models focused on the AGB phase of intermediate
mass stars ($M\geq 3$\msun) with metallicites 
$10^{-4} \leq Z \leq 6\times 10^{-4}$. These results complete previous 
investigations of higher Z AGB models by our group.
We find that low Z models, due to the lower amount of CNO present in their 
H-burning layers, attain larger temperatures both in the CNO burning shell 
and at the bottom of their surface convective region, thus achieving more 
easily than their higher Z counterparts HBB conditions. For a given luminosity,
they suffer a smaller mass loss (due the their smaller radii), thus the total 
number of thermal pulses that they experience is higher.
In the low mass regime ($M<5$\msun), where the effects of the third dredge-up 
dominate over hot bottom burning, the sum of C+N+O along with Na, Mg and
Al increase with decreasing Z (the Al production in these models is favoured by
the use of the upper limits of the cross sections of the proton capture
reactions by the magnesium isotopes). For masses $M>5$\msun~
HBB takes over as the main physical process changing the surface chemistry, 
and the abundances of those elements that at high temperatures suffer 
destruction by proton capture, primarily oxygen and sodium, diminish for 
decreasing Z; conversely, the content of carbon and nitrogen are seen to
be Z-independent for masses close to the limit for carbon ignition in
degenerate conditions.
A comparison of our yields with those by a different research group 
confirms that the treatment of convection plays the most striking role
in determining the essential evolutionary properties of these class of
objects. 

The O-Na anticorrelation shown by NGC~6397 stars can be explained if
dilution at a level of $\sim 50\%$ between the mass ejected by AGBs
and pristine matter is adopted; this is also in good agreement with the
resulting helium mass fraction of the SG stars, Y$\sim 0.28 - 0.30$, that is
consistent with the morphology of the HB of this cluster. \\
A comparison with recent spectroscopic results of CN abundances in low
luminosity M15 giants shows that the chemistry of the theoretical ejecta 
of our models agree with the most extreme chemistries observed, i.e. 
with the stars showing the largest nitrogen enhancement and carbon reduction.
This seems to confirm even at these low metallicities that self-enrichment 
by massive AGBs is responsible for the star-to-star differences observed,
though more observational results are needed before drawing more robust
conclusions. 

Very oxygen--poor stars ([O/Fe]$\sim -1$) are predicted to exist,
as SG stars formed by the winds of our most massive models of
$Z=2\times 10^{-4}$, at least in the most massive very low metallicity GCs.

\begin{acknowledgements}
The authors are grateful to the anonymous referee for his detailed and
careful reading of this work, that greatly helped improving the quality
of the manuscript
\end{acknowledgements}


\begin{thebibliography}{}
   
   \bibitem[1999]{angulo} Angulo, C., Arnould, M., Rayet, M. 
      et al. 1999, Nucl. Phys. A, 656, 3
	
	\bibitem[Bedin et al.(2004)]{bedin2004} Bedin, L.~R., Piotto, G.,
Anderson, J., Cassisi, S., King, I.~R., Momany, Y., Carraro, G. 2004,
ApJ, 605, L125
  
	\bibitem[Beers \& Sommer-Larsen(1995)]{beers1995} Beers, T.~C., \& Sommer-Larsen, J.\ 
	1995, ApJS, 96, 175 

\bibitem[Beers 
\& Christlieb(2005)]{beers2005} Beers, T.~C., \& Christlieb, N.\ 2005, \araa, 43, 531 
	  
   \bibitem[1995]{blocker2} Bl\"ocker, T. 1995, A\&A, 297, 727

   \bibitem[1991]{blocker1} Bl\"ocker, T., \& Sch\"onberner, D. 
      1991, A\&A, 244, L43

	\bibitem[Busso et al.(2007)]{busso2007} Busso, G., et al.\ 2007, A\&A, 474, 105   
	  
   \bibitem[2005]{caloi1} Caloi, V, \& D'Antona, F. 2005, A\&A, 463, 987

   \bibitem[2007]{caloi2} Caloi, V, \& D'Antona, F. 2007, A\&A, 463, 949

   \bibitem[1996]{canuto2} Canuto, V.M.C., \& Mazzitelli, I. 1991, ApJ, 370, 295

   \bibitem[2006]{carretta1} Carretta, E. 2006, AJ, 131, 1766

   \bibitem[2006]{carretta2} Carretta, E., Bragaglia, A., Gratton, R.G., 
     Leone, F., Recio-Blanco, A., \& Lucatello, S. 2006, A\&A, 450, 523

   \bibitem[1998]{cavallo} Cavallo, R.M., Sweigart, A.V., \& Bell, R.A. 
      1998, ApJ, 492, 575

   \bibitem[1976]{cloutman} Cloutman, L., \& Eoll, J.G. 
      1976, ApJ, 206, 548

   \bibitem[2005]{cohen} Cohen, J.G., Briley, M.M., \& Stetson, P.B. 
      2005, AJ, 130, 1177

   \bibitem[Cottrell \& Da Costa(1981)]{1981ApJ...245L..79C} Cottrell, P.~L., 
     \& Da Costa, G.~S.\ 1981, \apjl, 245, L79    	  
	  
   \bibitem[2002]{franca2} D'Antona, F., Caloi, V., Montalban, J., Ventura, P.,
     \& Gratton, R. 2002, A\&A, 395, 69 

   \bibitem[D'Antona \& Caloi(2004)]{franca1} D'Antona, F., \& Caloi, V. 2004, 
     ApJ, 611, 871

   \bibitem[D'Antona \& Caloi(2008)]{franca2008} D'Antona, F., \& Caloi, V. 2008, 
     MNRAS, in press
	 
   \bibitem[D'Antona et al.(2005)]{dantona2005} D'Antona, F., 
      Bellazzini, M., Caloi, V., Pecci, F.~F., Galleti, S., \& Rood, R.~T.\ 2005, 
      \apj, 631, 868 

   \bibitem[1996]{franca4} D'Antona, F., \& Mazzitelli, I. 1996, ApJ, 473, 550

   \bibitem[2007]{franca3} D'Antona, F., \& Ventura, P. 2007, MNRAS, 379, 1431

   \bibitem[2007]{Decressin} Decressin, T., Meynet, G., Charbonnel, C.,
      Prantzos, N, Ekstr\"om, S. 2007, A\&A, 464 1029

   \bibitem[1998]{pavel5} Denissenkov, P., Da Costa, G.S., Norris, J.E.,
      \& Weiss, A. 1998, A\&A, 333, 926

   \bibitem[2003]{pavel2} Denissenkov, P., \& Herwig, F. 
      2003, ApJ, 590, L99

   \bibitem[1996]{pavel3} Denissenkov, P., \& Weiss, A. 1996, A\&A, 308, 773

   \bibitem[2001]{pavel4} Denissenkov, P., \& Weiss, A. 2001, ApJ, 559, L115

   \bibitem[D'Ercole et al.(2008)]{dercole2008} D'Ercole, A., 
Vesperini, E., D'Antona, F., McMillan, S.~L.~W., 
\& Recchi, S.\ 2008, MNRAS, in press (arXiv:0809.1438) 

   \bibitem[2008]{eggleton} Eggleton, P.P., Deaborn, D.S.P., \&
      Lattanzio, J.C. 2008, ApJ, 677, 581

   \bibitem[2004]{fenner} Fenner, Y., Campbell, S., Karakas, A.I.,
      Lattanzio, J.C., \& Gibson, B.K. 2004, MNRAS, 353, 789

   \bibitem[2005]{ferguson} Ferguson, J.W., Alexander, D.R., Allard, F., 
      et al. 2005, ApJ, 623, 585

   \bibitem[2004]{formicola} Formicola, A., Imbriani, G., Costantini, H., 
      et al. 2004, Phys Lett. B, 591, 61

\bibitem[Frost et al.(1998)]{frost} Frost, C.~A., Cannon, R.~C., Lattanzio, J.~C., Wood, P.~R., \& Forestini, M.\ 1998, \aap, 332, L17 

\bibitem[2003]{harris2003} Harris, W.E. 2003, Catalog of parameters for the Milky Way Globular Clusters,
http://www.physics.mcmaster.ca/~harris/mwgc.dat

   \bibitem[2001]{gratton2} Gratton, R., Bonifacio, P., Bragaglia, A., et al. 
      2001, A\&A, 369, 87
      
 \bibitem[2003]{gratton2003} Gratton, R.~G., Bragaglia, A., Carretta, E., Clementini, G., Desidera, S., Grundahl, F., \& Lucatello, S.\ 2003, A\&A, 408, 529      

   \bibitem[2004]{gratton1} Gratton, R., Sneden, C., \& Carretta, E. 
      2004, ARA\&A, 42, 385
      
         \bibitem[1998]{GS98} Grevesse, N., \& Sauval, A.J. 1998, SSRv, 85, 161
  
   \bibitem[Herwig(2004)]{herwig2004} Herwig, F.\ 2004, \apjs, 155, 651 

   \bibitem[2005]{herwig1} Herwig, F. 2005, ARA\&A, 43, 435

   \bibitem[2002]{hale1} Hale, S.E., Champagne, A.E., Iliadis, C., 
      Hansper, V.Y., Powell, D.C., \& Blackmon, J.C. 2002, Phys.Rev.C, 65, 5801

   \bibitem[2004]{hale2} Hale, S.E., Champagne, A.E., Iliadis, C., 
      Hansper, V.Y., Powell, D.C., \& Blackmon, J.C. 2004, Phys.Rev.C, 70, 5802

   \bibitem[1996]{iglesias} Iglesias, C.A., \& Rogers, F.J. 
      1996, ApJ, 464, 943

   \bibitem[1999]{ines2} Ivans, I.I., Sneden, C., Kraft, R.P.,  
      et al. 1999, AJ, 118, 1273

   \bibitem[2007]{izzard} Izzard, R.G., Lugaro, M., Karakas, A.I., Iliadis, C., 
      \& van Raai, M. 2007, A\&A, 466, 641

\bibitem[Johnson et al.(2007)]{johnson2007} Johnson, J.~A., Herwig, 
F., Beers, T.~C., \& Christlieb, N.\ 2007, \apj, 658, 1203 

   \bibitem[2007]{amanda} Karakas, A., Lattanzio, J.C. 2007, PASA, 24, 103

   \bibitem[1994]{kraft} Kraft, R.P. 1994, PASP, 106, 553
   
   \bibitem[Kroupa et al.(1993)]{kroupa1993} Kroupa, P., Tout, C.~A., 
\& Gilmore, G.\ 1993, \mnras, 262, 545 

   \bibitem[Lee et al.(2005)]{lee2005} Lee, Y.-W., et al.\ 2005, 
ApJ Letters, 621, L57 

  \bibitem[Lucatello et al.(2005)]{2005ApJ...625..825L} Lucatello, S., 
Tsangarides, S., Beers, T.~C., Carretta, E., Gratton, R.~G., 
\& Ryan, S.~G.\ 2005, ApJ, 625, 825 
   
   
   \bibitem[Maiolino et al.(2004)]{maiolino2004} Maiolino, R., Oliva, E., Ghinassi, F., 
   Pedani, M., Mannucci, F., Mujica, R., \& Juarez, Y.\ 2004, A \& A, 420, 889 
   
   \bibitem[Norris(2004)]{norris2004} Norris, J.~E.\ 2004, ApJ Letters, 
612, L25 
   
   \bibitem[Piotto et al.(2005)]{piotto2005} Piotto, G., et al.\ 
2005, ApJ, 621, 777 

   \bibitem[Piotto et al.(2007)]{piotto2007} Piotto, G., Bedin, L.R., Anderson, J. 
       et al.\ 2007, \apj, 661, L53 

   \bibitem[Prantzos \& Charbonnel(2006)]{prantzos1} Prantzos, A., \& Charbonnel, C. 2006, A\&A, 458, 135
   
   \bibitem[Salpeter(1955)]{salpeter1955} Salpeter, E.~E.\ 1955, \apj, 
121, 161 

   \bibitem[1995]{SCV} Saumon, D., Chabrier, G., \& Van Horn, H.M. 1995, ApJS, 99, 713
   
   \bibitem[2005]{smith} Smith, V.V., Cunha, K., Ivans, I.I., et al. 2005,
      ApJ, 633, 392 

   \bibitem[2004]{sneden1} Sneden, C., Kraft, R.P., Guhathakurta, P., 
      Peterson, R.C., \& Fulbright, J.P. 2004, AJ, 127, 2162

   \bibitem[1991]{sneden2} Sneden, C., Kraft, R.P., Prosser, C.F., 
      \& Langer, G.E. 1991, AJ, 102, 2001

   \bibitem[1997]{sneden3} Sneden, C., Kraft, R.P., Shetrone, M.D., 
      Smith, G.H., Langer, G.E., \& Prosser, C.F. 1997, AJ, 114, 1964

   \bibitem[2000]{SB2000} Stolzmann, W., \& Bl\"oecker, T. 2000, A\&A, 361, 1152

   \bibitem[1979]{sweigart} Sweigart, A.V., \& Mengel, J.G. 1979, ApJ, 229, 624

\bibitem[2000]{Anthony-Twarog2000} Anthony-Twarog, B.~J., \& Twarog, B.~A.\ 2000, AJ, 120, 3111 

   \bibitem[van Loon et al.(1999)]{vanloon1999} van Loon, J.~T., Zijlstra, A.~A., 
      \& Groenewegen, M.~A.~T.\ 1999, \aap, 346, 805 

   \bibitem[2005a]{ventura2} Ventura, P., \& D'Antona, F.  
      2005a, A\&A, 431, 279

   \bibitem[2005b]{ventura3} Ventura, P., \& D'Antona, F. 
      2005b, A\&A, 439, 1075

   \bibitem[2006]{ventura7} Ventura, P., \& D'Antona, F. 
      2006, A\&A, 457, 995

   \bibitem[2008]{ventura9} Ventura, P., \& D'Antona, F. 
      2008a, A\&A, 479, 805

  \bibitem[Ventura \& D'Antona(2008)]{ventura2008MNRAS} Ventura, P., \& D'Antona, F.\ 
     2008b, MNRAS, 385, 2034 

   \bibitem[2000]{ventura8} Ventura, P., D'Antona, F., \& Mazzitelli, I. 
      2000, A\&A, 363, 605

   \bibitem[2002]{ventura1} Ventura, P., D'Antona, F., \& Mazzitelli, I. 
      2002, A\&A, 393, 215

   \bibitem[2001]{ventura5} Ventura, P., D'Antona, F., Mazzitelli, I., 
      \& Gratton, R. 2001, ApJ, 550, L65

   \bibitem[1998]{ventura6} Ventura, P., Zeppieri, A., D'Antona, F., 
      \& Mazzitelli, I., 1998, A\&A, 334, 953

   \bibitem[1953]{vitense} Vitense, E. 1953, Zs.Ap., 32, 135
   
   \bibitem[Wachter et al.(2008)]{wachter2008} Wachter, A., Winters, J.~M., 
    Schr{\"o}der, K.-P., \& Sedlmayr, E.\ 2008, \aap, 486, 497 

   
   
\end{thebibliography}
\end{document}